\documentclass[12pt,preprint]{aastex}

\def\simlt{\mathrel{\spose{\lower 3pt\hbox{$\mathchar''218$}}
     \raise 2.0pt\hbox{$\mathchar''13C$}}}
\def\simgt{\mathrel{\spose{\lower 3pt\hbox{$\mathchar''218$}}
     \raise 2.0pt\hbox{$\mathchar''13E$}}}
\begin{document}
\def\gtorder{\mathrel{\raise.3ex\hbox{$>$}\mkern-14mu
             \lower0.6ex\hbox{$\sim$}}}
\def\ltorder{\mathrel{\raise.3ex\hbox{$<$}\mkern-14mu
             \lower0.6ex\hbox{$\sim$}}}

\def\today{\number\year\space \ifcase\month\or  January\or February\or
        March\or April\or May\or June\or July\or August\or
        September\or
        October\or November\or December\fi\space \number\day}
\def\fraction#1/#2{\leavevmode\kern.1em
 \raise.5ex\hbox{\the\scriptfont0 #1}\kern-.1em
 /\kern-.15em\lower.25ex\hbox{\the\scriptfont0 #2}}
\def\spose#1{\hbox to 0pt{#1\hss}}
\def\heion{\ion{He}{2}}
\def\wig#1{\mathrel{\hbox{\hbox to 0pt{%
          \lower.5ex\hbox{$\sim$}\hss}\raise.4ex\hbox{$#1$}}}}
\def\Teff{T_{\rm eff}}
\def\sss{\scriptscriptstyle}

\title{Clouds and Chemistry: Ultracool Dwarf Atmospheric Properties from 
Optical and Infrared Colors}
\author{Mark S. Marley\footnote{NASA Ames Research Center, MS 245-5, 
Moffett Field CA, 94035},
S. Seager\footnote{Institute for Advanced Study, Einstein Drive, Princeton, NJ 08540}, 
D. Saumon\footnote{Department of Physics and Astronomy, Vanderbilt University, Nashville, TN 37235}, 
Katharina Lodders\footnote{Planetary Chemistry Laboratory, Department of Earth
and Planetary Sciences, Washington University, St. Louis, MO 63130-4899},
Andrew S. Ackerman$^1$, Richard S. Freedman$^1$, Xiaohui Fan$^2$}

\pagestyle{plain}

\begin{abstract}
The optical and infrared colors of L and T dwarfs are sensitive to
cloud sedimentation and chemical equilibrium processes in their
atmospheres.  The $i'-z'$ vs. $J-K$ color-color diagram provides a
window into diverse atmospheric processes mainly because 
different chemical processes govern each color, and cloud opacity largely
affects $J-K$ but not $i'-z'$.  Using theoretical atmosphere models
that include for the first time a self-consistent treatment of cloud
formation, we present an interpretation of the $i'-z'$ vs. $J-K$ color
trends
of known L and T dwarfs.  We find that the $i'-z'$ color is extremely
sensitive to chemical equilibrium assumptions: chemical equilibrium
models accounting for cloud sedimentation predict redder $i'-z'$
colors---by up to 2 magnitudes---than models that neglect
sedimentation.  We explore the previously known $J-K$ color trends
where objects first become redder, then bluer with decreasing effective
temperature. Only models that include sedimentation of condensates
are able to reproduce these trends.  We find that the exact
track of a
cooling brown dwarf in $J-K$ (and $i'-z'$) is very sensitive to the
details of clouds, in particular to the efficiency of sedimentation of
condensates in its atmosphere.  We also find that clouds still affect
the strength of the $J$, $H$, and $K$ band fluxes of even the coolest 
T dwarfs. In addition, we predict the
 locus in the $i'-z'$ vs.  $J-K$ color-color diagram of brown dwarfs
cooler than yet discovered.

\end{abstract}
\keywords{stars: atmospheres --- stars: low-mass, brown dwarfs}
\section{Introduction}
The Sloan Digital Sky Survey (SDSS) and the 2-Micron All Sky Survey
(2MASS) have both had great success in discovering L and T type ultracool
dwarfs. The colors of these objects provide insight into
the processes operating in their atmospheres. In the SDSS system, 
all such objects are uniquely red in $i'-z'$.  
L dwarfs are red in the 2MASS $J-K_s$ color $(1 \le J-K_s \le 2)$ 
while the cooler T
dwarfs like Gliese 229 B are bluer $(J-K_s < 0.5)$ (Kirkpatrick et
al. 1999; Burgasser et al. 1999; Burgasser et al. 2000).  While
the mechanisms responsible for the $J-K_s$ and the $i'-z'$ colors of
the L and T dwarfs are generally understood, there has yet been no single theory
that self-consistently describes the evolution of ultracool 
dwarfs\footnote{We reserve the term `brown dwarf' only for
the unmistakably substellar T-dwarfs.  `Ultracool dwarfs' encompasses
all objects later than and including the late M dwarfs.  
Late M and early L objects may or may not be substellar.}
in this color space.

Because of their intrinsic faintness, moderate to high resolution
spectroscopy may not be performed on all of the 
ultracool dwarfs discovered by these surveys.  Thus analyses of
ultracool dwarf colors could be essential to provide information on their
physical properties. We have explored the
utility of a number of $i'$, $z'$, $J$, $H$, and $K$ color-color
diagrams for constraining physical properties and find the
$i'-z'$ vs. $J-K$ diagram particularly interesting.  First, these are
essentially the colors in which SDSS and 2MASS discover ultracool
dwarfs\footnote{Note that 2MASS employs the $K_s$ filter in its
survey.  Henceforth we exclusively employ $K$ band.}.
Second, alkali
metals dominate the $i'-z'$ colors while H$_2$O and CH$_4$ absorption
bands and cloud physics control the $J-K$ colors.  Over the pressure
and temperature ranges of interest, the chemical pathways of alkali metals
and H$_2$O/CH$_4$/CO are not strongly coupled, thus this particular color-color
diagram reflects a remarkably diverse set of chemical effects.

In this paper we describe how clouds and the chemistry of carbon,
oxygen and alkali elements (mainly potassium) control the 2MASS and SDSS
colors used to discover ultracool dwarfs, and we explore the potential of
the $i'-z'$ vs. $J-K$ color-color diagram as a tool to deduce the
physical characteristics of dwarfs and the physics
of their atmospheres.  We also predict the colors of
very cool brown dwarfs, those with effective temperatures $\Teff \wig< 700\,$K,
which are yet to be discovered.

\section{Color Trends}
\label{sec:colortrends}
Ultracool dwarfs are notoriously different from blackbodies of the same
effective temperature.  Figure~\ref{fig:colors} shows the $i'-z'$ vs. $J-K$
colors for observed SDSS L and T dwarfs 
(Strauss et al. 1999; Fan et al. 2000; Leggett et al. 2000; 
Tsvetanov et al. 2000; Geballe et al. 2001a).  The ultracool dwarfs are
spread out over several magnitudes in both $i'-z'$ and $J-K$. In
addition, they are located in very different parts of this diagram.

Blackbodies become redder in all colors with decreasing $T_{\rm eff}$ as
the Planck peak shifts redward; a temperature sequence of blackbody
emitters in Figure~\ref{fig:colors} would follow a diagonal line
cutting from blue to red through the extreme upper left corner of the
color-color diagram. Ultracool dwarfs, however, are subject to a more
complex set of influences and first become redder and then bluer in
$J-K$ as they age and cool.  The initial reddening arises as
progressively larger amounts of condensates are found in the visible 
atmospheres in
the  $T_{\rm eff}$ range from $\sim 2000$ to $\sim 1400\,\rm K$.  At
lower effective temperatures
$J-K$ turns blueward because in the cooler brown dwarfs the cloud base
(and thus most of the cloud opacity) falls below the photosphere
(Marley 2000; Ackerman \& Marley 2001; Tsuji 2001; Allard et
al. 2001), leaving the visible atmosphere relatively clear of
condensates. In the absence of clouds, opacities such as water, methane, and
pressure-induced absorption by molecular hydrogen
act to rapidly close the $K$ band infrared window as $\Teff$ falls,
resulting in increasingly blue $J-K$.

In the optical, known ultracool dwarfs become redder with decreasing $T_{\rm
eff}$. This trend is produced by
 the growing importance of the $0.59\,\mu$m Na I and $0.77\,\mu$m K I 
resonance doublets (Tsuji, Ohnaka \& Aoki 1999; Burrows, Marley,
\& Sharp 2000) with decreasing $T_{\rm eff}$; as the dwarf cools,
the gradual disappearance of TiO and
cloud opacity leaves a progressively more transparent
atmosphere at optical wavelengths.  The K I resonance doublet is centered on the
$i'$ band while the $z'$ band is only affected by the far red wing, producing very 
red $i'-z'$ colors (Fig.~\ref{fig:filter}).  We predict below that this
trend should reverse in objects with lower $T_{\rm eff}$ than have yet been
observed.

Brown dwarfs with $\Teff$ and infrared colors intermediate between the coolest
and reddest L dwarfs and the much cooler and bluer T dwarfs like Gl
229 B were initially thought to be rare since 2MASS found few of them
(Kirkpatrick et al. 2000). 
T dwarfs with $1100 \wig< \Teff \wig< 1300\,$K are difficult to discover
in the 2MASS $J-K_s$ color because their colors are similar to the far 
more numerous and hotter M dwarfs.
The SDSS optical colors do not suffer from this infrared color degeneracy in 
this $T_{\rm eff}$ range.  The SDSS
collaboration found the first brown dwarfs with
$J-K$ colors lying between 0.5 and 1 (formerly called L/T transition
objects; see Leggett et al. 2000) and has now typed them as early T
dwarfs (Geballe et al. 2001a).  

\section{Model Atmospheres}
To model the colors of solar metallicity L and T dwarfs we employ the
radiative-convective equilibrium atmosphere model of Marley et
al. (1996; further described in Burrows et al. 1997).  The model has
been updated to self-consistently include both alkali opacities as 
described in Burrows et
al.  (2000) and the precipating clouds of Ackerman \& Marley
(2001).  The treatment of the clouds and the chemistry is described
further below.

High resolution spectra are computed from these atmosphere
models (temperature profile and cloud structure) with a spectral
synthesis code (Saumon et al.  2000; Geballe et al. 2001b).  In the
high-resolution spectra, the non-isotropic scattering by dust
particles is mapped onto an equivalent isotropic scattering problem
following the prescription given in Chamberlain \& Hunten (1987).
Theoretical colors in turn are calculated from the high resolution
synthetic spectra.  For $J$ and $K$ colors we use the
Mauna Kea Observatory (MKO) Near Infrared
System (Simons \& Tokunaga 2001; Tokunaga, Simons \& Vacca 2001)
and for SDSS the $i'$ and $z'$ filter functions 
and the AB magnitude system (Fukugita et al. 1996).

\subsection{Cloud Model}
   For the radiative transfer calculations the clouds are assumed to 
be horizontally
homogeneous and are modeled following the approach developed by 
Ackerman \& Marley
(2001).  This approach assumes a steady state in which the upward 
transport of gas
and condensate by turbulent mixing is balanced by the downward transport of
condensate by sedimentation.  In convective regions the turbulent mixing in the
model is determined by the convective heat flux, and in radiative 
regions the mixing
is determined by a minimum eddy diffusion coefficient, a prescribed 
parameter that
characterizes such processes as breaking buoyancy waves.  The sedimentation
(or precipitation) in the model is determined by the condensate mass,
the convective velocity, and $f_{\rm rain}$, an 
adjustable parameter
that describes the efficiency of sedimentation relative to the 
turbulent mixing.
Physically, $f_{\rm rain}$ represents the combined effects of unresolved dynamical and
microphysical details such as the skewness of atmospheric circulations and the
abundance of condensation nuclei.  Larger values of $f_{\rm rain}$ correspond 
to greater
precipitation and hence thinner clouds.  Note that the base of a cloud is fixed
at the lowest level where the partial pressure of a condensible exceeds its
saturation vapor pressure.  Hence, any precipitation that falls 
through the base
of a cloud is assumed to evaporate, returning its mass to the 
reservoir of vapor
below the cloud.  Precipitation through cloud base does not 
annihilate the cloud;
instead the (steady-state) cloud is continually replenished by condensation in
updrafts from below, as in long-lived terrestrial clouds.

   The value of $f_{\rm rain}$ and the profile of turbulent mixing together 
determine the
profile of condensate mass in the clouds; by assuming that the cloud particle sizes
follow a lognormal distribution in a manner consistent with the turbulent 
mixing and
sedimentation, the model also calculates a vertical profile of cloud opacity.
Ackerman \& Marley (2001) find that their model best fits the observations of
condensate scale height, particle size, and optical depth in Jupiter's ammonia
cloud deck with a value of $f_{\rm rain} = 3$.

The cloud structure and atmosphere temperature profiles are solved to
converge simultaneously and self-consistently by the atmosphere
code.   As the atmospheric temperature structure as a function of
pressure, $T(P)$, is adjusted by the
convergence algorithm, a new cloud profile is computed following
Ackerman \& Marley (2001). In the course of the calculation of a
single temperature-pressure profile for a specified $T_{\rm eff}$ and
gravity $g$, many hundreds of trial $T(P)$ and associated cloud profiles 
are computed.  An atmospheric structure is not considered
acceptable unless both the temperature structure and the cloud model
have simultaneously and self-consistently converged
\footnote{Ackerman \& Marley (2001) presented model cloud profiles 
computed from several fixed temperature profiles to illustrate the
characteristics of their cloud algorithm.  In the current paper
we use the Ackerman \& Marley algorithm iteratively to solve for a
self-consistent atmospheric profile. The statement in Ackerman \&
Marley (2001) that the cloud profiles are not self consistent with
the atmosphere model does not apply to the current contribution.}.

In this work we include only Fe, $\rm MgSiO_3$ (representing both 
$\rm Mg_2SiO_4$ and $\rm MgSiO_3$), and $\rm H_2O$ as condensates.  
Other species (e.g. $\rm Al_2O_3$)
either condense below the optically thick Fe cloud or are relatively
insignificant opacity sources (see Marley 2000).  For example, 
in the $T_{\rm eff} = 2000\,\rm K$ model the $\rm Al_2O_3$ cloud falls
in a region of the atmosphere that is already opaque.  The additional
opacity arising from the cloud does not alter the adiabatic
temperature profile.  The
overlying silicate and iron clouds play a far more important role.
For hotter cases where silicates do not condense, $\rm Al_2O_3$ 
is more important.

Figure~\ref{fig:profiles}  presents several of our temperature-pressure
profiles for $T_{\rm eff}=2000$ and 1300 K. For each $T_{\rm eff}$ 
a cloud free and two cloudy models are shown.
Our cloud free models are computed with the same set of
assumptions for
chemical equilibrium as are our cloudy models (condensed species are 
segregated by settling and no longer interact with the gas), 
but with all cloud opacity removed to isolate the effect of the
clouds (see \S3.2 for comparison
with the models of Allard et al. (2001)).  Condensation equilibrium
curves establish the cloud base level for
each profile.  Two $T_{\rm eff}=1300\,\rm K$ profiles from Tsuji (2001)
are also shown.

For the $T_{\rm eff}=2000\,\rm K$ models
silicate and iron grains form above the radiative-convective
boundary and their influence on the radiative temperature profile
is apparent.  The cloudy models are substantially warmer than the
equivalent cloud free case.  As expected, the (optically
and physically) thicker $f_{\rm rain}=3$ 
cloud produces an even  greater
thermal perturbation than the case with more efficient sedimentation ($f_{\rm rain}=5$).
 
In the case of the 1300 K cloudy models, the cloud base is located
within the convective region.  The temperature profile within this region
is set by the adiabatic lapse rate.  Since the cloud
simply adds to the (already high) opacity and the thermal profile is
controlled by the adiabatic lapse rate, perturbations along
the atmospheric thermal profile comparable to the hotter case are not seen.  
The clouds do play a role in raising the top of the convection zone above 
what it would be in the otherwise identical cloud free case by
adding opacity above the cloud-free radiative-convective boundary.
Above the cloudy radiative-convective boundary the cloud-top opacity 
is sufficient
to keep the radiative portions of the atmosphere warmer than in the cloud-free
case.  

The entropy at the radiative-convective boundary 
controls the adiabat upon which
the deeper atmosphere---and consequently the entire interior of the
ultracool dwarf---resides (Burrows et al. 1997). The cloudier the upper
atmosphere (smaller $f_{\rm rain}$), the hotter the interior.
The interior structure at a fixed effective temperature and the
amount of energy which must be radiated away to cool the entire
dwarf to a lower effective temperature are thus affected by even
small differences in cloud opacity.
Hence different cloud structure
assumptions produce different cooling histories. We plan to explore
such effects in a future work.

Different assumptions regarding the cloud models result in very different
thermal profiles.  For example the cloud free model from 
Tsuji (2001) shown  in Figure~\ref{fig:profiles}
is quite similar to our own result for the same $g$ and $T_{\rm eff}$.
Also shown is a model from Tsuji (2001) (Tsuji's case B) in which there is no 
removal of condensates from the atmosphere above the cloud base.
In this case the greenhouse heating of the atmosphere by the abundant
dust far exceeds what our cloudy models with sedimentation predict.
The upper atmosphere in this $T_{\rm eff} = 1300\,\rm K$ case reaches
temperatures as high as those found in our cloudiest 
$T_{\rm eff}=2000\,\rm K$ case.   This example dramatically highlights
the important role sedimentation plays in moderating what would otherwise
be an overpowering role of dust  in controlling the temperature-pressure
profile of the atmosphere.  Chabrier et al. (2000) discuss the
dissociation of water occuring in the atmospheres of their hot, dusty 
no-sedimentation cases (their `DUSTY' models).  The large dissociation
fractions in those models are simply driven by the lack of any sedimentation 
and thus represent particularly extreme--and likely unphysical--cases.
Although not shown in Figure~\ref{fig:profiles} for the sake of clarity,
Tsuji (2000) also presents a `unified' model in which the top of the
cloud is simply terminated at an arbitrary temperature.
Such a model produces little to no heating in the
atmosphere above the cloud top and substantial heating below the cloud
top (comparable to our $f_{\rm rain} = 5$ case for $T_{\rm eff} = 1300\,\rm K$).
Ultimately only detailed fitting of observed spectra and colors will distinguish
between all such possibilities.

\subsection{Chemical Equilibrium Model}

The calculation of chemical equilibrium in an atmosphere is dependent
upon the assumptions made regarding the fate of condensates.  In a
gravitational field, atmospheric constituents that condense tend to
fall.   If the condensate
is liquid water meteorologists term it rain.  We consider two different
chemical equilibrium models. In the first case there is no
sedimentation of condensates.  For this we use the baseline model
from Burrows and Sharp (1999; hereafter BS99).  In the second case we
treat sedimentation with the cloud condensation model developed by
Lewis (1969) for Jovian planets and used by Fegley and Lodders (1996),
Lodders (1999), and Lodders and Fegley (2001) for brown dwarfs.

Note that
there is a slight inconsistency between the vertical distribution of
condensates in the chemical equilibrium model (using the vertical
profile described by Lewis (1969)) and the radiative
transfer cloud model (using the 
model of Ackerman \& Marley (2001)).  However, the
vertical condensate profiles with moderate values of $f_{\rm rain}$
are similar to those predicted by the Lewis model.  See Ackerman \&
Marley (2001) for more details.

For the purposes of comparison, we have also computed a sequence of
cloud free models.  In these models, the presence of condensates is
taken into account in the calculation of the chemical equilibrium
but the opacity of condensates is ignored in the calculation of radiative
transfer.  These models differ from the `COND' models of Allard
et al. (2001).  In the Allard et al. models the chemical equilibrium always assumes
the presence of grains even if they are not included in the
radiative transfer.

\section{The Optical-IR Color-Color Diagram}
Figure~\ref{fig:colors} shows a temperature sequence of ultracool dwarf models
in the $i'-z'$ vs. $J-K$ color-color diagram.  Models are plotted for
a fixed surface gravity of $1000$~m~s$^{-2}$, corresponding
roughly to a mass of 35 Jupiter masses ($\rm M_J$).  Note that the
surface gravity, $g$, of a given object increases as it contracts and
cools, so for a given object the cooling track will follow a
slightly different path. Evolution paths for ultracool dwarfs of different masses, however, 
are almost degenerate in the color-color
diagram because the temperature at which optical depth 2/3
is reached as a function of wavelength depends only weakly on the gravity. 
All surface gravities very nearly
overlap in the $i'-z'$ vs. $J-K$ color-color diagram.  So although $T_{\rm eff}$
may be estimated, there is no unique ($T_{\rm eff}, g$) solution for
given $i'-z'$ and $J-K$ colors.

The $i'-z'$ vs. $J-K$ color-color diagram is very sensitive to $T_{\rm
eff}$ because of the disparate chemistry governing the two colors.
The alkali metal chemistry for the observed ultracool dwarfs shown in
Figure~\ref{fig:colors} mostly consists of neutral K being depleted
into molecules and solids. This process (see Lodders 1999 for a
complete discussion) is not strongly coupled to the C/H/O chemistry
that controls CO, CH$_4$ and H$_2$O partitioning. At even lower
temperatures, K disappears into chloride and hydroxide gases but
the alkali chemistry is still only weakly coupled to the C, H,
and O chemistry. As a result dwarfs at different $T_{\rm eff}$ are
well separated in this color-color diagram. There is no degeneracy
for different $T_{\rm eff}$ as found in most other color-color
combinations (e.g. $H-K_s$ vs. $J-H$, $J-K_s$ vs. $I-J$ (see
Kirkpatrick et al. 2000 and Tsuji 2001)).

\subsection{Clouds}
The behavior of a cloud layer as a function of $\Teff$ is of primary
astrophysical interest.  Qualitatively, the base of the cloud occurs
where the $(T,P)$ structure of the atmosphere crosses the condensation
curve of the major condensates (silicate and iron at high
temperatures, and water at lower temperatures).  A cloud deck forms
with a vertical profile determined by the cloud model.
Because in the region of interest the condensation
temperature of relevant substances increases weakly with pressure,
the base of the cloud layer occurs at a nearly constant (but slowly increasing)
temperature as $\Teff$ decreases.  On the other hand, the
temperature of the photosphere is approximately $\Teff$.  It follows
that as $\Teff$ decreases, the cloud layer gradually disappears below
the observable level of the atmosphere.  This phenomenon has been
discussed by several authors (Chabrier et al. 2000; Marley 2000;
Allard et al. 2001; Tsuji 2001).

The opacity of the gas in ultracool dwarfs is dominated by molecular bands
and varies strongly with wavelength. In contrast Mie scattering by
large particles produces a nearly grey cloud opacity.
Thus the above discussion is somewhat
simplistic since the concept of photosphere is not well defined in
these objects.  While continuum opacities ensure that the photosphere
corresponds approximately to a fixed physical level in normal stars,
in brown dwarfs the visible and near-infrared spectrum can probe a
range of depths of up to 6 pressure scale heights (Saumon et
al. 2000).  This range provides an opportunity to observationally probe the
vertical structure of brown dwarf atmospheres.

The gradual disappearance of the cloud layer below the
``photosphere'' as $\Teff$ decreases is illustrated in
Figure~\ref{fig:clouddepth} where the curves show the level in the
atmosphere where the optical depth $\tau_{\sss \nu}=2/3$.  Here,
vertical position in the atmosphere is indicated by the local
temperature.  Three cases are shown with $\Teff=500$, 1000, and
1500$\,$K from top to bottom, respectively.  A pair of curves is shown
for each model; one showing the photosphere (where optical depth
$2/3$ is reached) determined by gas opacity
only and one for the nearly grey cloud opacity  only.  For
the upper pair of curves ($\Teff=500\,$K), the deep silicate and iron 
cloud ``photosphere'' lies
below (at higher temperature) the gas photosphere at all wavelengths,
implying that the cloud layer remains essentially invisible and has
little effect on the emergent spectrum.  In the lower pair of curves
($\Teff=1500\,$K), the cloud becomes opaque well above the gas
photosphere in the $J$, $H$, and $K$ bands.  The cloud layer is
therefore observable in these three bandpasses (but not at other
wavelengths) and the spectral energy distribution is strongly affected
by the presence of the cloud.

Figure~\ref{fig:clouddepth} clearly shows that the cloud layer disappears
below the observable atmosphere over a range of effective temperatures, 
depending on the
bandpass of observation.  For example, the cloud becomes invisible in the
$K$ band for $\Teff \wig< 1400\,$K but remains detectable in the $J$ band
down to $\Teff \sim 800\,$K.  The Ackerman \& Marley cloud model with 
$f_{\rm rain}=5$ implies that the spectra of all known T dwarfs are affected 
by clouds.

\begin{figure}
\epsscale{0.8}
\plottwo{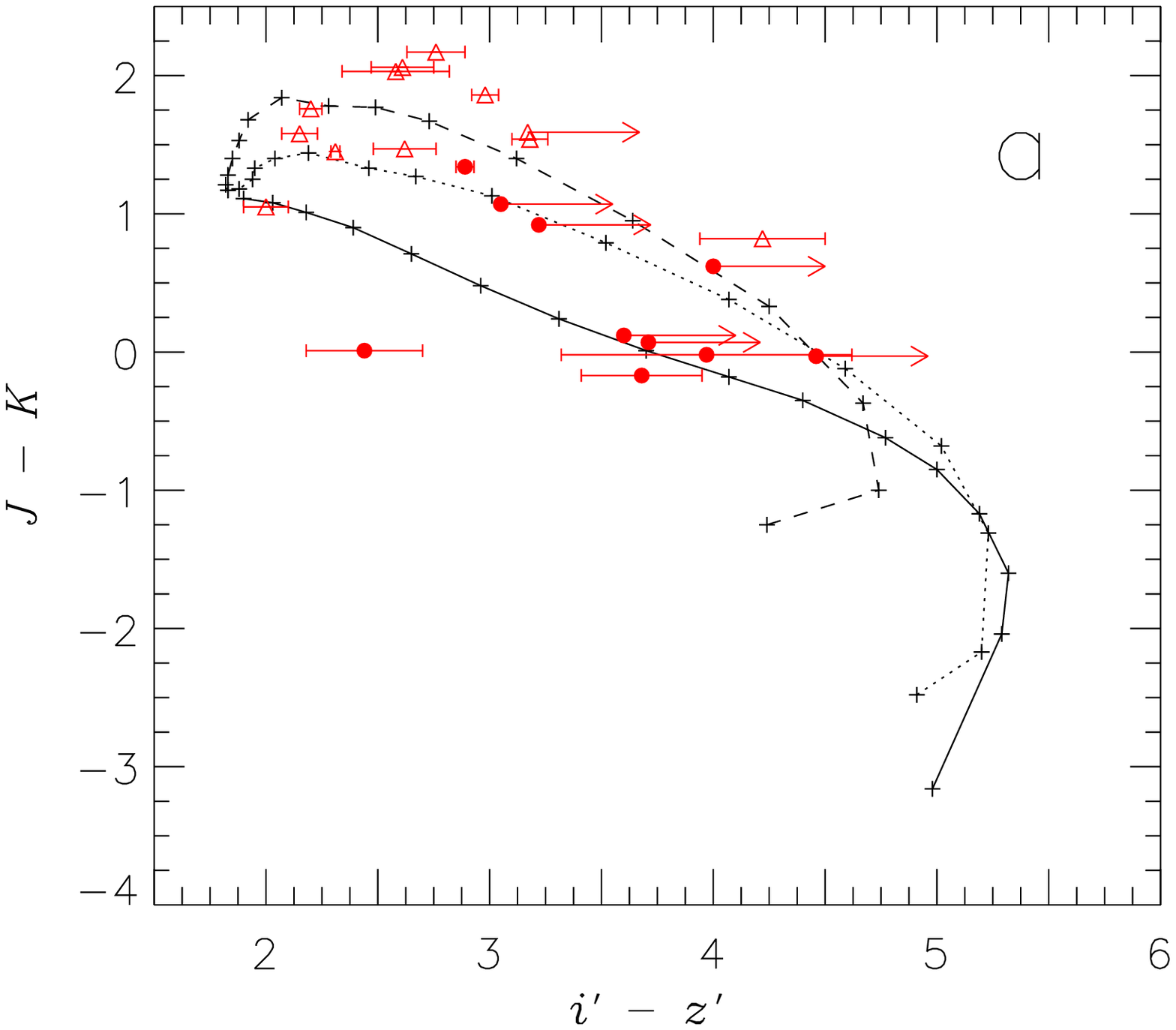}{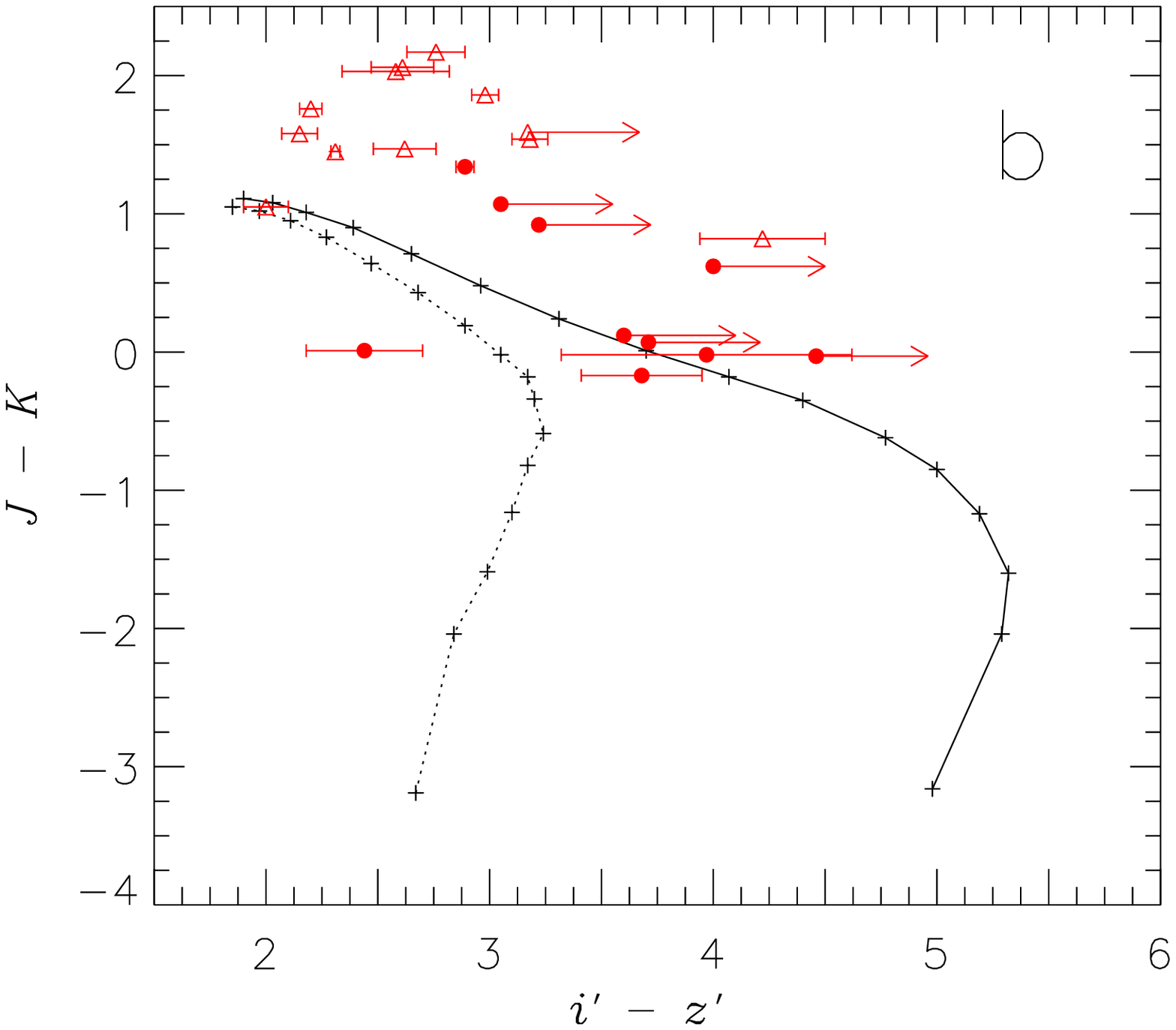}
\caption{$i'-z'$ (SDSS) vs. $J-K$ (MKO system) color-color diagrams for
L (triangles) and T (circles)
dwarfs (Strauss et al. 1999; Fan et al. 2000; Legett 
et al. 2000; Tsvetanvo et al. 2000; Leggett et al. 2001; Geballe
et al. 2001a).  Plotted SDSS
magnitudes have been converted to the AB system while the 
MKO magnitudes are in the Vega system.  Lower limits are denoted
by arrows.  The lines show our theoretical models
computed in the same systems with the symbols representing $T_{\rm eff}$ 
steps of 100$\,$K in the
$T_{\rm eff}$ range 2000$\,$K -- 400$\,$K.  Panel (a) shows the results of
cloudy models (dashed and dotted lines with $f_{\rm rain}=3$
and 5, respectively) and cloud free models (solid line) for 
$g = 1000\,\rm m\,s^{-2}$.  Panel (b) shows cloud free models with 
$g = 1000\,\rm m\,s^{-2}$. The solid line shows models with the
sedimentation chemical equilibrium model by Lodders and the dotted
line are models using the BS99 chemical equilibrium model with no
condensate sedimentation. See \S3 and \S4 for details.  The anomalous
data point at $i'-z'=2.44$, $J-K=0.01$ represents SDSSJ020742.83+000056.2.
\label{fig:colors}}
\end{figure}

Observationally, one of the most revealing features in the $i'-z'$
vs. $J-K$ color-color diagram shown in Figure~\ref{fig:colors} is the
reddening in $J-K$ of the L dwarfs that is not present in the T dwarfs.
This difference in $J-K$ trajectory
results from the presence of condensates throughout the photosphere of the L
dwarfs but not in the late T dwarfs. 
The blackbody-like condensate emission pushes L
dwarfs to the red in $J-K$, despite the tug of water opacity towards
the blue.  This effect of condensate opacity
is best illustrated by comparing the cloud free models and
the cloudy models 
in Figure~\ref{fig:colors}a. The cloud free L dwarf models
show a continuous blueward trend in $J-K$ with decreasing $T_{\rm
eff}$ --- because of increasing H$_2$O and pressure-induced $\rm H_2$ 
absorption --- in contradiction with the redward trend of the L dwarf
data. The cloudy models on the other hand, generally match the redward trend in
$J-K$ of the L dwarf data.

The progressively redder $J-K$ colors of L dwarfs
has been noted before (e.g. Kirkpatrick et al. 1999; Martin et
al. 1999; Fan et al. 2000; Leggett et al. 2001; Tsuji 2001) 
and demonstrated by spectral
fitting to be caused by the appearance of more and more silicate
condensates in the cooling ultracool dwarf atmospheres (e.g Leggett,
Allard, \& Hauschildt 1998; Burrows, Marley \& Sharp 2000; 
Chabrier et al. 2000; Marley 2000).
However models in which there is no settling of the condensates
(Chabrier et al. 2000) produce colors, particularly for the later
L dwarfs, that are much too red.  For example the dusty model of
Chabrier et al. predicts that a 1 Gyr old $50\,\rm M_J$ brown dwarf
with $T_{\rm eff} = 1424\,\rm K$ will have $J-K = 3.9$.  In fact
the reddest L-dwarfs have $J-K \approx 2.2$ (see Fig. 4 of
Leggett et al. 2001). Our models
with $f_{\rm rain} = 3$ peak at $J-K \sim 1.8$ for
$T_{\rm eff} = 1400\,\rm K$.  The muted $J-K$ colors of the reddest L dwarfs
provide strong evidence of condensate sedimentation.

A second revealing feature in the $i'-z'$ vs. $J-K$ color-color
diagram is the transition between the L and early T dwarfs that
begins as a blueward turnover in $J-K$ in the latest
L dwarfs (Leggett et al. 2001).  As the condensates sink below
the visible atmosphere, their blackbody effect is
removed, halting the redward $J-K$ progression. As molecular
opacities ($\rm H_2$, $\rm H_2O$, and later $\rm CH_4$) become predominant,
their greater absorption at $K$ band initiates the turn in $J-K$ to the blue.
This turnover occurs when 
the cloud layer is no longer visible in the $K$ band (see Fig.~\ref{fig:clouddepth}).  
An important issue has been the
temperature range over which the L to T transition occurs (e.g. Reid et
al. 2001).  The model $T_{\rm eff}$ at which the turnover begins, as well as
the maximum value of $J-K$, depend on the sedimentation parameter
$f_{\rm rain}$.  Of the models shown in
Figure~\ref{fig:colors}, $f_{\rm rain} = 3$ 
comes closest to matching the observed turnover in $J-K$.  
Smaller values of $f_{\rm rain}$ produce thicker,
more massive clouds and somewhat lower values may better fit the peak 
$J-K$ at the turnover.  The cloud tops remain in view in the $J$ and $K$ bands
down to cooler $\Teff$.  This $J-K$ blueward
turnover likely will be better characterized by future SDSS discoveries, and the
data will be essential for understanding cloud properties in ultracool
dwarfs.

At some lower $T_{\rm eff}$ ($ \sim 800\,$~K for $f_{\rm rain}=5$) the
base of the condensate cloud base is below the
visible photosphere.  However, the tops of the
silicate clouds might still be limiting the depths from which flux
emerges in the water and methane windows, thus accounting for the
difficulty all cloud free models have had in correctly reproducing the
ratio of the flux emerging from within and without of the water bands
(Allard et al. 1996; Marley et al. 1996; Tsuji et al. 1996; Saumon et
al.  2000; Geballe et al. 2001b).

\subsection{Alkali Metal Chemistry}
\begin{figure}
\plotone{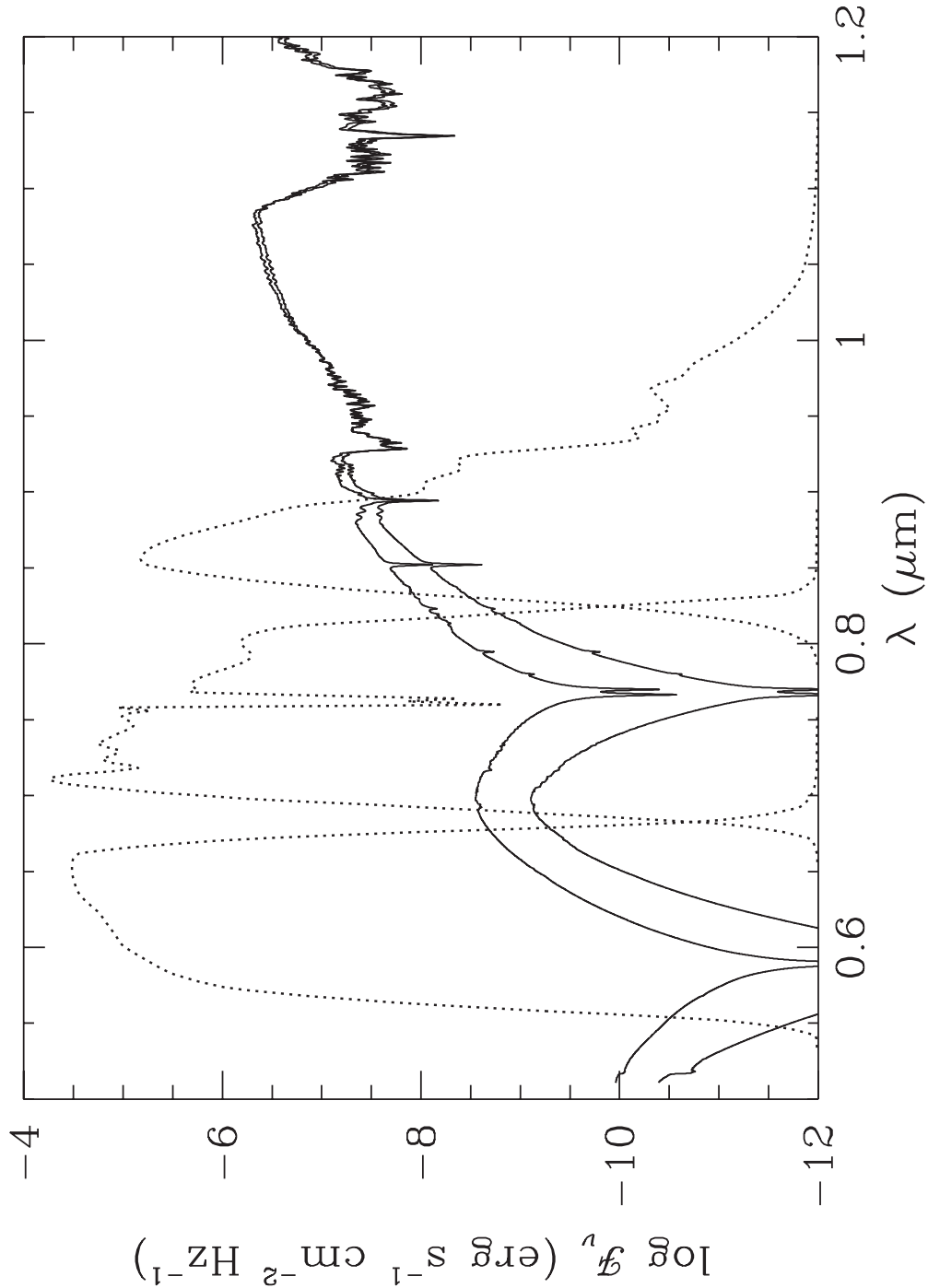}
\caption{SDSS $r'$, $i'$, and $z'$ transmission curves (dotted lines
from left to right, respectively) superimposed on brown dwarf model
spectra. The $i'$ flux is controlled by the K I doublet line core and
the $z'$ flux by the K I doublet wing.  These cloud free synthetic
spectra with $T_{\rm eff}=1000\,$K and $g=1000\,\rm m\,s^{-2}$ are computed
with the chemical abundances of BS99 (i.e. no sedimentation assumed;
upper curve) and of Lodders (with condensate sedimentation; lower curve).
The effect of the chemical equilibrium model on the strength of the K I and Na I
doublets is very noticeable.  See \S3.2 and \S4.2 for details.
\label{fig:filter}}
\end{figure}

\begin{figure}
\epsscale{0.8}
\plotone{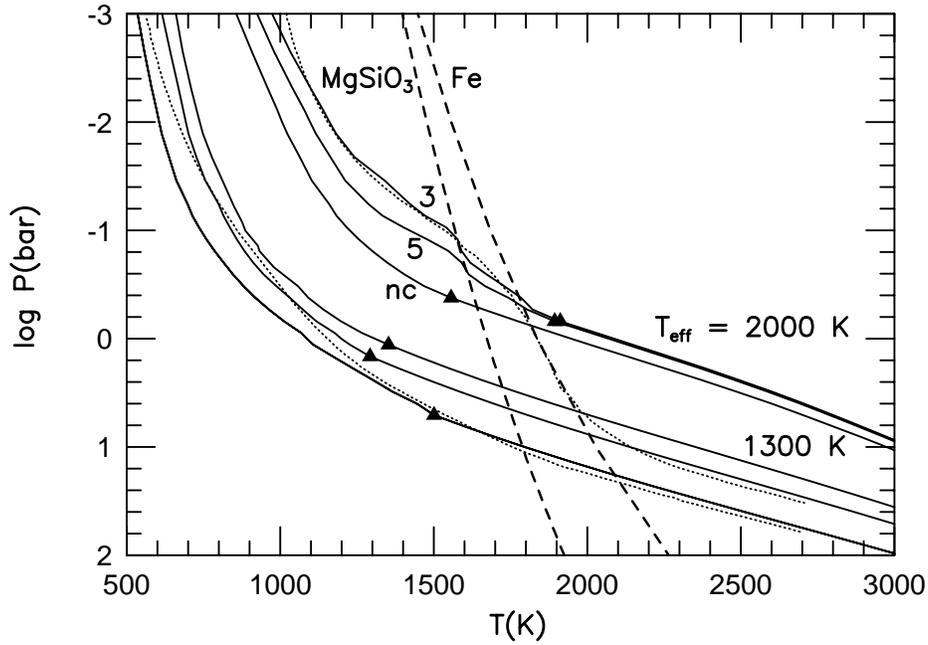}
\caption{Radiative-convective equilibrium atmosphere models for
$g=1000\,\rm m\,s^{-2}$.  For two values of $T_{\rm eff}$ solid lines
illustrate self-consistent temperature profiles calculated for the case
of no cloud opacity (nc) and rainfall efficiency factor $f_{\rm rain}=3$
and 5 (Ackerman \& Marley 2001). Lines are labelled for the 2000 K case.
For 1300 K the sequence of curves is the same.  Triangles denote 
convective-radiative zone boundaries; 
the deepest region of the atmosphere is always convective.  Dotted lines
show 1300 K models by Tsuji (2001) without condensation (left curve) and
with condensation but no sedimentation (right curve).  Dashed lines show the
condensation curves of enstatite and iron.
\label{fig:profiles}}
\end{figure}

\begin{figure}
\epsscale{0.9}
\plotone{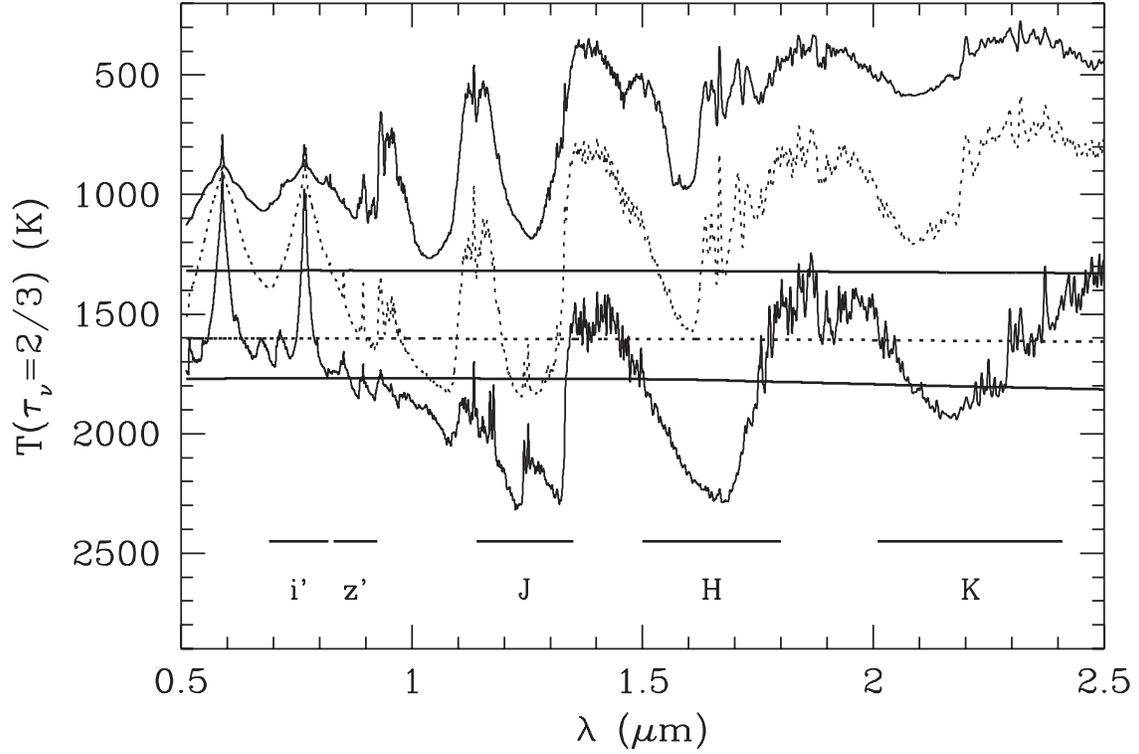}
\caption{Visibility of the cloud layer in brown dwarfs as a function
of $\Teff$.  The curves show the depth of the photosphere ($\tau_{\sss
\nu}=2/3$), indicated by the temperature in the atmosphere, as a
function of wavelength.  The abcissa is essentially a brightness
temperature.  Three models with $g=1000\,$m$\,$s$^{-2}$ and $f_{\rm
rain}=5$ are shown, from top to bottom $\Teff=500$, 1000 (dotted), and
1500$\,$K, respectively.  Two curves are shown for each model, one
showing the photosphere due to gas opacity only, and one due to cloud
opacity only.  The latter is very flat, due to the nearly grey cloud
opacity, and shows the level where the cloud becomes optically thick.
At wavelengths longer than shown here, the cloud remains below the
photosphere for all models.  Bandpasses for several filters are
indicated along the bottom of the figure.
\label{fig:clouddepth}}
\end{figure}

The $i'$ and $z'$ band fluxes are diagnostic of alkali metal
chemistry, mainly because they measure the core and the wing of the K
I resonance doublet, respectively. In T dwarfs, the red wing of the
doublet is detected up to 200~nm from the line core (Burrows et
al. 2000).  Figure~\ref{fig:filter} shows the $i'$ and $z'$ filters
superimposed on two different model spectra.  The $i'$ filter is
centered on the K I doublet core and the $z'$ filter probes the far
red wing. The ultracool dwarf colors become redder in $i'-z'$ for
decreasing $T_{\rm eff}$ because these filters probe the Wien tail of the 
Planck function and the K I doublet gets stronger.  The
gradual disappearance of TiO and cloud opacity as $T_{\rm eff}$
decreases leaves behind a nearly transparent atmosphere at
wavelengths below $1\, \mu$m (Figure~\ref{fig:clouddepth}) and reveals the
K I doublet in all its pressure-broadened splendor.
At low $T_{\rm eff}$
($\sim 700\,$K) the $i'-z'$ redward trend halts as K I is depleted
into KCl and the doublet weakens.

Given the dependence of the $i'-z'$ color on the K I resonance
doublet, this color provides a stringent test for chemical equilibrium
models.  The two curves in Figure~\ref{fig:colors}b show
colors computed with and without the assumption of condensate sedimentation in 
the chemical equilibrium calculation.  There is a substantial 
difference --- of 2
magnitudes --- in $i'-z'$ at effective temperatures where the K I line
is prominent ($\sim 800\,\rm K$).  The major difference between the
two approaches is that at temperatures below 1400 K, the monatomic K
abundance (hence the opacity) is greatly reduced under the assumption of no 
sedimentation (BS99) compared to the assumption of sedimentation 
(Lodders 1999). A comparison of spectra
computed under both assumptions is shown in Figure~\ref{fig:filter}.
The effect on the $i'-z'$ color is rather dramatic and the models
without sedimentation turn blueward well before the model that
includes sedimentation, as shown in Figure~\ref{fig:colors}.

Because the SDSS T-dwarfs are only marginally detected in $i'$ band, 
errorbars for those
objects shown in Figure~\ref{fig:colors} are substantial.  
The trends in T-dwarf data
shown in the figure are generally closer to the sedimentation chemistry
models, but more and better $i'$-band detections are required to
fully support this conclusion.

The two different assumptions used to model chemical equilibrium of
gas and condensates give such different results that they are worth
discussing in more detail. The two models depend on the physical
setting (see Lodders 1999; Lodders \& Fegley 2001). In the
no-sedimentation case condensates remain in local equilibrium with the
gas. In cooler regions, the high temperature (primary)
condensates react with the upper atmospheric gas to form secondary
condensates via gas-solid reactions. Complete chemical equilibrium
exists between all phases in this no-sedimentation case. BS99 term
this case the ``no rainout" approach.  Their approach (also employed
by e.g. Chabrier et al. 2000; Allard et al. 2001)
implies that alkali elements such as Na and K
condense into alkali feldspar ((Na,K)AlSi$_3$O$_8$)
after a long sequence of primary condensate reactions with the gas.
The net effect in this no-sedimentation case is that the gaseous
atomic K and Na become depleted in the atmosphere once alkali feldspar
condenses.
 
As described in detail by Lodders (1999) and Lodders \& Fegley (2001),
however, this approach does not apply to ultracool dwarf and giant planet
atmospheres because a gravity field is acting on condensates. The
primary condensates are sequestered by sedimentation into a cloud and
are not available for gas-solid reactions in the atmosphere above the
cloud layer as the dwarf cools.  These cloud condensation models
have been used successfully for over 30 years in the planetary
community (Lewis 1969, Barshay \& Lewis 1978, Fegley \& Lodders 1994)
and were recently termed ``rainout" by BS99.  We prefer to use
the term `sedimentation' because `rainout' could be interpretted as implying
complete removal.  In the
sedimentation case, elements such as Al and Ca condense at
greater depth and are consequently absent in the overlying
atmosphere.  Thus alkali feldspar cannot form, and
Na and K remain in the gas phase.  Only when a brown dwarf is
much older and cooler ($T_{\rm eff}\sim 700\,\rm K$) are atmospheric
temperatures low enough for monatomic Na and K  to
convert into chloride and hydroxide gases.  At even lower
temperatures Na and K condense into $\rm Na_2S$ and KCl (see also
the discussions in Lodders (1999) and Burrows et al. (2000)).
 
The $i'-z'$ color is sensitive to pressure broadening of the K I doublet.
The exceptionally strong pressure broadening affecting
the $0.59\,\mu$m Na I and the $0.77\,\mu$m K I resonance doublets
in T dwarfs stretches the current theories of line broadening beyond
their limit of validity.  These lines are modeled with a far wing
exponential cutoff $\exp-(qh\Delta\nu/kT)$ where $q$ is an
undetermined parameter of order unity\footnote{The parameter $q$ may be measured
experimentally in the near future (A. Dalgarno, private communication).}.
A detailed discussion is given in Burrows et al. (2000), as are fits
of the optical spectra of Gl 229B and SDSS 1624+00.  With abundances
determined from the sedimentation chemistry of Lodders (1999), we have
obtained good fits of the optical spectra of Gl 229B and Gl 570D with
$q=1$ (Geballe et al. 2001b).  The $i'-z'$ color
changes by as much as 0.4 mag for models computed with $q=0.5$ and $q=1$.  
Disentangling the line broadening
parameters from other $i'-z'$ color effects will likely come from
fitting high resolution spectra.

\subsection{The Coolest Brown Dwarfs}

The coolest brown dwarf known with a reliable determination of 
its effective temperature is Gl 570D with $\Teff\sim800\,$K 
(Geballe et al. 2001b).
Cooler brown dwarfs  are expected to enter a new regime in the $i'-z'$ vs.
$J-K$ color space than those discovered so far. Brown
dwarfs with $\Teff \ltorder 600\,$~K are expected to have water clouds
forming in the upper atmosphere. Just as the subsidence of silicate clouds below 
the photosphere causes a turnover in colors, the
appearance of water clouds in the upper reaches of low-$\Teff$ atmospheres
could have dramatic effects on the colors.

At a relatively cool effective temperature ($\sim 600\,\rm K$), as K I disappears into KCl,
the $i'-z'$ color reaches a maximum and turns 
blueward as suggested by the coolest objects in Figure~\ref{fig:colors}.
The formation of significant ($\tau > 0.1$) water clouds 
below $T_{\rm eff}\sim 500\,$K (depending on $g$)
halts and may eventually reverse the blueward march in 
$J-K$ with decreasing $T_{\rm eff}$
because the water cloud acts like a blackbody, redistributing the flux to
the blackbody peak.   The models presented here may underestimate
the size of the redward turn in $J-K$.  Smaller particles
than the $\sim 20$ to $30\,\rm \mu m$ predicted by the cloud model
would arise for smaller values of the unconstrained stratospheric
eddy diffusion coefficient and would produce more cloud opacity 
for $T_{\rm eff} \le
500 \,\rm K$.  Such objects will be very faint at $z'$ and
will be difficult to detect with SDSS. Nevertheless, the number
density of brown dwarfs suggests that a few such objects
could be detected by SIRTF (Mart{\'\i}n et al. 2001).

\section{Discussion}
The $i'-z'$ vs. $J-K$ color-color diagram reveals the importance of
precipitating condensation clouds in controlling the colors
of the L dwarfs and the transition between L and T dwarfs, and will
complement high resolution spectroscopy (Griffith \& Yelle 2000; Geballe et al. 2001a) to
reveal the nature of condensation chemistry in these atmospheres.

Most previous and current ultracool dwarf models (e.g. 
Allard et al. 1996; Marley et al. 1996; Tsuji et al. 1996; Burrows et al. 1997; 
Chabrier et al. 2000; Allard et al. 2001) considered either the case in 
which condensates
remain suspended in the atmosphere or considered them to be absent
from the photosphere due to sedimentation. In contrast Marley (2000) and
Tsuji (2001) considered cloud decks confined to some
fraction of a pressure scale height.  The Marley and Tsuji models, although
including no cloud physics, were both able to produce a
red to blue transition in $J-K$.  By including for the first time a
self-consistent treatment of cloud physics, we demonstrate that sedimentation
processes in clouds result in model $J-K$ colors that are much
less red---by up to 2.5 magnitudes---than models with no
sedimentation (Chabrier et al. 2000).  Sedimentation controls
the cloud vertical extent and is responsible for the observed turnover in 
$J-K$ with decreasing effective temperature.
The model further predicts that
the spectra and colors of even the coolest known T dwarfs are influenced
by clouds.  

Furthermore our atmosphere model is the first to compute particle
sizes simultaneously and self-consistently with the thermal profile.
Both Allard et al. (2001) and Tsuji (2001) assume a fixed, submicron, particle size
distribution derived
from interstellar medium dust grains. Allard et al. correctly point
out that as long as the particle size is smaller than the wavelength
of light, Rayleigh scattering dominates the opacities and the exact
size distribution of particles has little effect on the opacities.
They also argue that particle sizes larger than 100~$\mu$m are
implausible because they would break up (terrestrial raindrops and
billiard-ball sized hailstones bely this assertion).  Our model
predicts silicate and iron particle sizes between 10--$100\,\mu$m.  
Such large particles are Mie, not Rayleigh, scatterers in the
near-infrared and possess a completely different spectral opacity (see
Figure 3 in Marley (2000)) than the submicron particles assumed
by other groups.

The models, however, do not provide a perfect fit to the available data.
As noted in section 4.1, the peak model $J-K$ (1.8) is not quite as red as
the peak observed value (2.2).  L dwarfs with the largest $J-K$ 
range in $i'-z'$ 
from 2.5 to 3.0.  At the $J-K$ peak the $f_{\rm rain} = 3$ model
predicts $i'-z'=2.1$.    This discrepancy may arise from the large uncertainty
in the alkali pressure broadening.
The value of $f_{\rm rain}$ which comes closest to matching the peak in 
$J-K$ (observationally an L5 or L6 object, Leggett et al. 2001) does so at a
model effective temperature of $1400\,\rm K$.  This is slightly cooler
than the range expected for such an object (see Burgasser et al. 2001).
Of greater concern is that this model then moves too slowly to the
blue.  The earliest T dwarfs have $J-K \sim 1.3$ (Leggett et al.
2001).  The $f_{\rm rain}=3$ model reaches this point at $T_{\rm eff} =
1000\,\rm K$ which is certainly too cool.  Hence it appears that
different values of $f_{\rm rain}$ are required for the early
to mid L dwarfs (a Jupiter-like $f_{\rm rain}\sim 3$) and the latest L's and the
T dwarfs ($f_{\rm rain}\sim 5$ or larger)
Alternatively Ackerman \& Marley
(2001) have suggested that holes might preferentially appear in the 
cloud decks of
later type L dwarfs as the clouds begin to form within the convective region
of the atmosphere.  Bright, relatively blue cloud-free flux 
emerging from the holes
may help hasten the L to T transition in $J-K$.
If this is the case a complete description of the disk-averaged emitted
flux would by necessity include both relatively cloudy and clear regions.
 
The optical $i'-z'$
color is strongly affected by the presence of monatomic potassium and
modeling this color relies on the treatment of the alkali condensation
chemistry. Chemical equilibrium models not accounting for sedimentation of
condensates result in lower KI abundances because potassium is removed from
the gas by alkali-feldspar at higher temperatures. Hence the
no-sedimentation models yield up to two magnitudes bluer $i'-z'$ colors than
models where sedimentation of condensates is taken into account. This is
because the sedimentation of high temperature condensates prevents alkali
feldspar from forming and K I abundances are higher until monatomic K is
converted into KCl gas and KCl condensation sets in at lower temperatures.
Improved brown dwarf $i'-z'$ colors will reveal which treatment of the
equilibrium chemistry in brown dwarf atmospheres is correct.  Since
the best-fitting cloud model predicts that cloud particles are not
lofted much above the cloud base, the sedimentation chemistry is
likely most appropriate, in agreement with physically based expectations.  A complete
test of this hypothesis, however, requires more accurate photometry
since brown dwarfs are usually not detected in $i'$ band by the SDSS
survey.  The follow up photometry is in progress.

It is now clear that the interpretation of objects from the warmest L dwarfs to the
the coolest T dwarfs requires an understanding of cloud formation in
ultracool dwarf atmospheres.  
Indeed more complex models, motivated perhaps by time resolved photometry and spectroscopy, 
will be needed to address many fundamental issues.
There is no question that what some have termed the field of `astrometeorology'
is still in its infancy.

\acknowledgements We thank Adam Burrows for use of the results from
his chemical equilibrium model and alkali gas opacity table.  We thank
the anonymous reviewers who provided many helpful comments which
markedly improved the presentation of this paper.
M.M. acknowledges support from NASA grants NAG2-6007 and NAG5-8919
and  NSF grants AST-9624878 and AST-0086288.  S.S. is supported by the 
W.M. Keck Foundation, and work by
K.L. and D.S. is supported by NSF grant AST-0086487 and NASA grant
NAG54988, respectively.

\end{document}